\documentclass[amssymb,useAMS,prd,aps,amsmath,superscriptaddress,nofootinbib,twocolumn]{revtex4}

\usepackage{color}
\usepackage{pslatex}
\usepackage{pdfpages}
\usepackage{graphicx}
\usepackage{psfrag}
\usepackage{pdftricks}
\usepackage{multirow}
\usepackage{graphicx}
\usepackage{hyperref}

\begin{document}

\title{The XENON100 exclusion limit without considering ${\cal{L}}_{\rm{eff}}$ as a nuisance parameter}

\author{Jonathan H Davis}
\affiliation{Institute for Particle Physics Phenomenology, University of Durham, Durham, DH1 3LE, UK}

\author{C\'eline B\oe hm}
\affiliation{Institute for Particle Physics Phenomenology, University of Durham, Durham, DH1 3LE, UK}
\affiliation{LAPTH, U. de Savoie, CNRS,  BP 110,
  74941 Annecy-Le-Vieux, France.}

\author{Niels Oppermann}
\affiliation{Max-Planck-Institut fur Astrophysik Karl-Schwarzschild-Str. 1. Postfach 13 17 85741 Garching, DE}

\author{Torsten Ensslin}
\affiliation{Max-Planck-Institut fur Astrophysik Karl-Schwarzschild-Str. 1. Postfach 13 17 85741 Garching, DE}

\author{Thomas Lacroix}
\affiliation{Ecole normale sup\'erieure de Lyon - Site Jacques Monod
46 all\'ee d'Italie. 69364 LYON cedex 07, France}

\date{today}

\begin{abstract}
In 2011, the XENON100 experiment has set unprecedented constraints on dark matter-nucleon interactions, excluding dark matter candidates with masses down to 6 GeV if the corresponding cross section is larger than  $10^{-39} \rm{cm^2}$. The dependence of the exclusion limit in terms of the scintillation efficiency (${\cal{L}}_{\rm{eff}}$) has been debated at length. 
To overcome possible criticisms XENON100 performed an analysis in which ${\cal{L}}_{\rm{eff}}$ was considered as a nuisance parameter and its uncertainties were profiled out by using a Gaussian likelihood in which the mean value corresponds to the best fit ${\cal{L}}_{\rm{eff}}$ value (smoothly extrapolated to zero below 3 keVnr). Although such a method seems fairly robust, it does not account for more extreme types of extrapolation nor does it enable to anticipate on how much the exclusion limit would vary if new data were to support a flat behaviour for ${\cal{L}}_{\rm{eff}}$ below 3 keVnr, for example. Yet, such a question is crucial for light dark matter models which are close to the published XENON100 limit. To answer this issue, we use a maximum Likelihood ratio analysis, as done by the XENON100 collaboration, but do not consider ${\cal{L}}_{\rm{eff}}$ as a nuisance parameter. Instead, ${\cal{L}}_{\rm{eff}}$ is obtained directly from the fits to the data. This enables us to define frequentist confidence intervals by marginalising over ${\cal{L}}_{\rm{eff}}$.
\end{abstract}

\maketitle

\section{Introduction}
After several decades of intensive searches,  direct detection experiments have now reached the required sensitivity to probe efficiently the parameter space associated with massive weakly interacting massive particles (WIMPs). Just a few years after the world's best exclusion limits set by the EDELWEISS and CDMS experiments on the dark matter-nucleon elastic scattering cross section, the XENON100 experiment demonstrated that the use of Xenon based technologies could actually beat Germanium detectors. By pushing down the exclusion limit by almost a factor 10 on the whole dark matter (DM) mass range (and with the present level of Krypton purity \cite{Aprile2011}), the XENON100 experiment not only could rule out values of the DM-nucleon cross section as low as $7 \ 10^{-45}$ cm$^2$ for DM particle masses of $\sim$ 50 GeV at 90$\%$ confidence level \cite{Aprile2011} but could also exclude (similarly to CDMS) some of the light DM candidates which have been hypothesised to explain CoGeNT \cite{Aalseth:2010vx}, DAMA/LIBRA \cite{Bernabei:2010mq} , CRESST \cite{Angloher:2011uu} and COUPP \cite{Behnke:2010xt}  findings (unless one relaxes some assumptions as done in e.g. \cite{Frandsen:2011ts,McCabe:2011sr,Kelso:2011gd,Kopp:2011yr} even though this may not be sufficient, e.g. \cite{Frandsen:2011gi}).

One element of controversy in the interpretation of these results is the dependence of this limit on the scintillation efficiency of the detector (see for example \cite{Collar2010,Savage2011,Collar2011}). A different ${\cal{L}}_{\rm{eff}}$ energy dependence at low nuclear recoil energy could indeed change the recoil energy associated with low mass WIMPs and possibly lead to a different exclusion limit than published in \cite{Aprile2011}. To address this issue, the XENON100 collaboration used a profile likelihood analysis in which ${\cal{L}}_{\rm{eff}}$ was taken to be a nuisance parameter and its uncertainties were profiled out 
with a Gaussian Likelihood $$ {\cal{L}}_{\rm{eff}}(t) = e^{-(t-t_{obs})^2/2}$$ where $t_{obs}=0$ represents the mean value of ${\cal{L}}_{\rm{eff}}$ smoothly extrapolated to zero (${\cal{L}}_{\rm{eff}}=0$) at low energy and $abs(t-t_{obs}) = 1$, the $1 \, \sigma$ confidence region. A flat behaviour of ${\cal{L}}_{\rm{eff}}$ at low energy would then be the upper limit of the 1-sigma contour.In principle, armed with such a modelling, the collaboration accounts for uncertainties in the extrapolation of ${\cal{L}}_{\rm{eff}}$ at low energy. This is important since the ${\cal{L}}_{\rm{eff}}$ energy behaviour below 3 keVnr is being currently debated.  

However, as the XENON100 analysis is very complex and relies on many sources of uncertainties, the use of this method makes it difficult to determine what would be the real impact 
of a very different energy behaviour of ${\cal{L}}_{\rm{eff}}$ (with respect to the mean considered by the XENON100 collaboration) on the exclusion limit if the latter was determined from new data with unprecedented precision below 3 keVnr (or if some existing data were found to be less reliable than previously thought). This is particularly important in the context of light dark matter candidates (eg. \cite{Boehm:2002yz}) where small recoil energies are expected.

Our work is essentially motivated by the fact that some light dark matter scenarios lie very close to the XENON100 limit and should be therefore very sensitive to new measurements of ${\cal{L}}_{\rm{eff}}$ below 3 keVnr. Several of these scenarios cannot be constrained by the LHC nor the measurement of the $\gamma$-ray, synchrotron fluxes nor even directional detectors (see for example \cite{Vasquez:2010ru,AlbornozVasquez:2011yq,AlbornozVasquez:2011js,Vasquez:2011bh}). Nuclear recoil direct detection experiments might be the only option to exclude or discover such scenarios, thus emphasising the need for a better determination of ${\cal{L}}_{\rm{eff}}$ at low energy. In order to properly account for the lack of determination of the low energy behaviour of ${\cal{L}}_{\rm{eff}}$, it is necessary to make transparent the correspondence between the ${\cal{L}}_{\rm{eff}}$ energy behaviour and the exclusion limit. Therefore, here, we will not treat ${\cal{L}}_{\rm{eff}}$ as a nuisance parameter. Instead we will use directly the value of ${\cal{L}}_{\rm{eff}}$ obtained from spline fits to the data and will not profile out the uncertainties (due in particular to the extrapolation at low energy) so as to quantify the effect ${\cal{L}}_{\rm{eff}}$ has on the exclusion limit. 

The estimate of the uncertainties presented in this work is limited since we are not part of the collaboration. It is likely that we do not use up-to-date methods and data (as a matter of fact new data should be published soon). Also we base our analysis on several assumptions which are explained in the sections below. However, despite all these limitations, we could recover the exclusion limit that XENON100 
collaboration has obtained and can therefore highlight the large impact the low energy behaviour of ${\cal{L}}_{\rm{eff}}$ has on the exclusion limit. Note that in this paper we focus on ${\cal{L}}_{\rm{eff}}$ uncertainties only; the astrophysical uncertainties will be addressed elsewhere.

In Section \ref{Leff}, we recall the spline interpolation to ${\cal{L}}_{\rm{eff}}$ dataset as well as the extrapolation at low energies and discuss the robustness of the fit using an extended filter formalism. We use different types of interpolation and extrapolation (consistent with the 1-sigma contour defined by XENON100 in \cite{Aprile2011}). In Section \ref{exclu}, we derive the exclusion limit for the mean ${\cal{L}}_{\rm{eff}}$ interpolation and compute the exclusion limits for more extreme ${\cal{L}}_{\rm{eff}}$ behaviour at low energies. Results are given in \ref{sec:results} and conclusion in \ref{sec:conclusion}.

\section{${\cal{L}}_{\rm{eff}}$ \label{Leff}}
The XENON100 experiment aims at detecting dark matter particles via their elastic scattering interactions with Xenon nuclei in a two-phase  
(liquid and gas) time-projection chamber (TPC) detector. A DM signal is then expected to have two signatures. The first one, referred to as the primary scintillation signal $S_1$, arises directly from the interaction of a DM particle with the liquid Xenon and measure the scintillation light in the liquid detector. The second, 
referred to as $S_2$, happens in the upper part of the detector -- at the liquid-gas interface -- and measures the scintillation light which results from the drift of the free electrons that originate from the ionisation of the Xenon nuclei in the liquid phase after the DM interaction and which survived the recombination with ionised atoms. 
Both signals are measured in photon-electrons units (PE) \cite{Aprile2011b} and are used to calibrate the detector's response to nuclear recoil events  and ultimately to determine whether the experiment has actually detected dark matter events. 

The discrimination parameter is defined as $$ \log \left(\frac{S_2}{S_1}\right) - ER_{\rm{mean}}.$$ 
Events below the threshold of $ER =   -0.4$ in the expected energy range are considered as potential DM events.
The XENON100 experiment uses the ratio of the two signals $S_1$ and $S_2$ to discriminate between a DM and a background event so the identification of signal is actually sensitive to the primary scintillation yield of recoiling Xenon nuclei in the liquid part of the detector. As the measurement of the absolute scintillation yield is difficult, the quantity that is used by the collaboration is the scintillation yield of nuclear recoils relative to that of 122 keV $\gamma$ rays from a $^{60}$Co source. This is called the relative scintillation efficiency and is referred to as ${\cal{L}}_{\rm{eff}}$. 

The nuclear-recoil energy threshold $E_{nr}$ (in units of keVnr) of a signal is then determined by both $S_1$ and ${\cal{L}}_{\rm{eff}}$ according to the relation, 
$$E_{nr} = \frac{S_1}{L_y \ {\cal{L}}_{\rm{eff}}} \frac{S_{er}}{S_{nr}},$$ where $L_y$  is a normalisation factor for the light-yield of the 122 keV gamma rays and $S_e$, $S_n$ are scintillation quenching factors for electronic and nuclear recoil respectively, due to the presence of an electric field (for XENON100, the values used are $S_e = 0.58$ and $S_n = 0.95$). 

The determination of $L_y$ and ${\cal{L}}_{\rm{eff}}$ are 
therefore of utmost importance.  While $L_y$ has been measured precisely to $L_y = 2.20 \pm 0.09 \rm{\frac{PE}{keVee}}$, there is no theoretical prediction for the energy dependence of ${\cal{L}}_{\rm{eff}}$. An empirical formula was obtained in \cite{Manzur2010,Mei2008} by fitting the data obtained in the same reference, namely $$ {\cal{L}}_{\rm{eff}} = q_{ncl} \ q_{esc} \ q_{el} $$ with $q_{ncl}$ the Lindhard factor (cf \cite{Manzur2010}), $q_{esc}$  reduction of the scintillation light yield and $q_{el}$ a quench factor due to bi-excitonic collisions \cite{Hitachi:2005ti}.

Such an empirical fit reproduces the observation that the ${\cal{L}}_{\rm{eff}}$ data decrease with decreasing energy and is also the assumption made by the XENON100 collaboration in \cite{Aprile2011} in order to obtain a conservative exclusion limit. Yet there are no measurement of ${\cal{L}}_{\rm{eff}}$ at low recoil energy. Besides, theoretical considerations by \cite{Bezrukov2011} seem to favour a constant behaviour of ${\cal{L}}_{\rm{eff}}$ at low energy. This would be consistent with the fit obtained by the XENON10 collaboration \cite{Angle:2011th}.  

The XENON100 collaboration's strategy to incorporate the uncertainties on ${\cal{L}}_{\rm{eff}}$ is to consider ${\cal{L}}_{\rm{eff}}$ as a nuisance parameter and profile out the uncertainties with a Gaussian Likelihood centred on the mean value of ${\cal{L}}_{\rm{eff}}$, that is the best fit.  Similar assumptions are made for the other parameters which enter the analysis. Although this seems a robust approach, it is not very transparent. In particular, one loses the correspondence between the exclusion limit and the uncertainties on ${\cal{L}}_{\rm{eff}}$ which arise due to a specific spline interpolation of the data and extrapolation at low energies. Indeed, the 1-sigma contour for ${\cal{L}}_{\rm{eff}}$ does not show on the exclusion curve obtained in \cite{Aprile2011} but since these uncertainties are due to the lack of data, one does expect to be able to keep track of them. In addition, it is hard to tell whether the final exclusion curve does take into account possible changes in the knots of the interpolation.

In the following, we therefore adopt a different strategy. We still use a profile Likelihood analysis but we do not treat ${\cal{L}}_{\rm{eff}}$ as a nuisance parameter. As a result, we can directly see the effect of the uncertainties on ${\cal{L}}_{\rm{eff}}$ interpolation and extrapolation on the exclusion limit. We thus obtain several exclusion limits where the mean should be seen as the exclusion limit corresponding to the best fit of ${\cal{L}}_{\rm{eff}}$ and where the edges of the contours correspond to the upper and lower parts of the ${\cal{L}}_{\rm{eff}}$ 1-sigma bands. Said differently, instead of obtaining one exclusion curve which would correspond to the best fit given all the uncertainties in the analysis, we prefer to draw the exclusion curves corresponding to the mean value and 1-sigma bands of ${\cal{L}}_{\rm{eff}}$ and let the reader marginalise 'by eyes' the effect of ${\cal{L}}_{\rm{eff}}$ on the exclusion curve. This approach enables us to anticipate the effect of a possible change in the physics of ${\cal{L}}_{\rm{eff}}$ below 3 keVnr.

\subsection{${\cal{L}}_{\rm{eff}}$ interpolation}

To overcome the lack of knowledge about the low energy behaviour of ${\cal{L}}_{\rm{eff}}$, it was suggested by the XENON100 collaboration to perform an interpolation of the Chepel et al. \cite{Chepel2006}, Manzur et al. \cite{Manzur2010}, Plante et al. \cite{Plante2011a} and Aprile et al. \cite{Aprile2009} ${\cal{L}}_{\rm{eff}}$ data sets and perform an extrapolation below 3 keVnr.  Since older data sets (e.g. \cite{Aprile2005}, \cite{Akimov2002},\cite{Arneodo:2000vc},\cite{Bernabei:2001pz}) were disregarded in \cite{Aprile2011}, we will only consider them to understand their impact on the ${\cal{L}}_{\rm{eff}}$ interpolation\footnote{An attempt was made by \cite{Sorensen2010} to measure ${\cal{L}}_{\rm{eff}}$ using the nuclear-recoil band of XENON10. This data is not considered in our fits, but does provide an interesting alternative method of determining the relative scintillation efficiency of Xenon.}. 

\begin{figure}[h]
\centering
\includegraphics[scale=0.4]{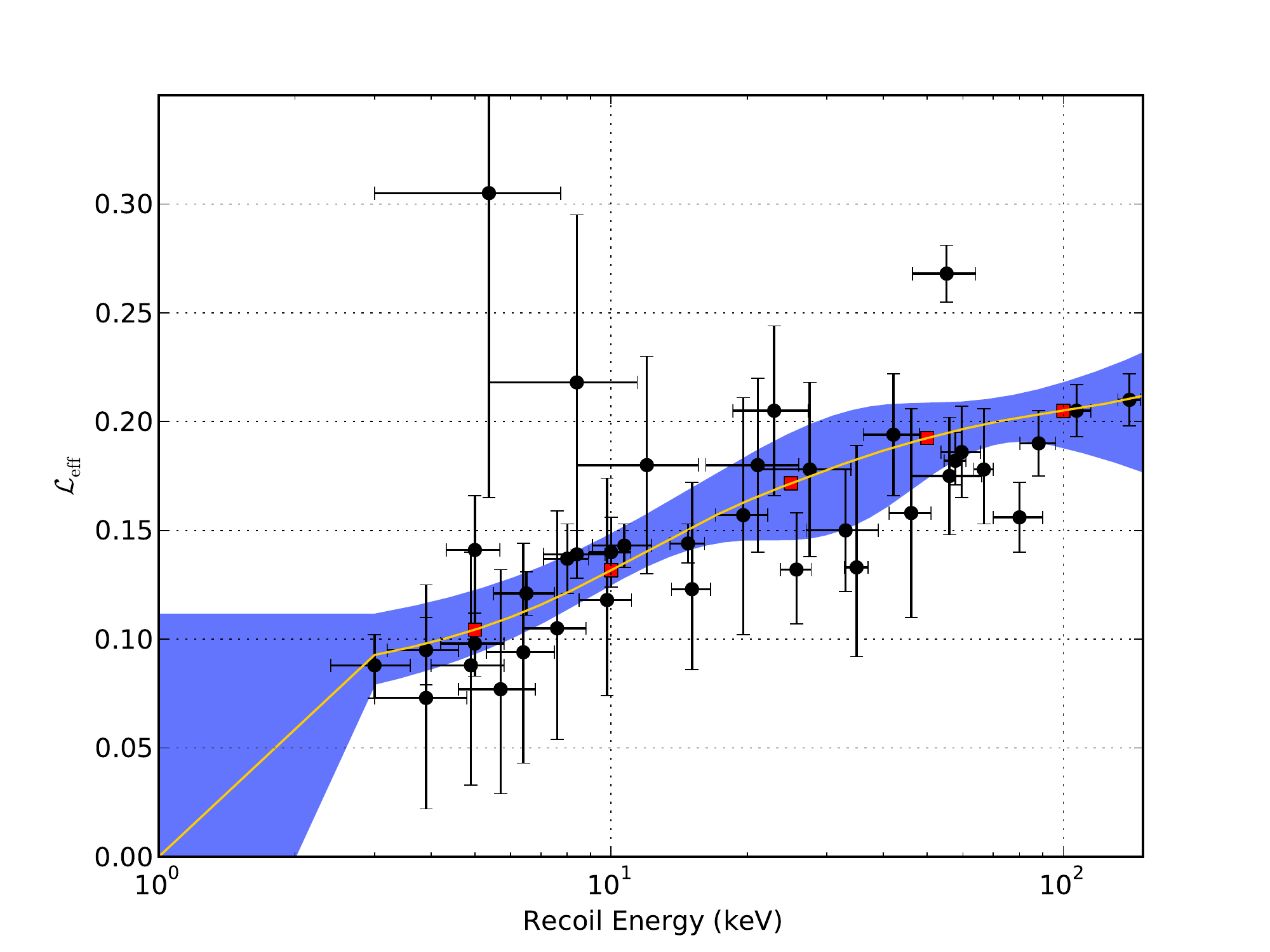}
\caption{A fit of a natural cubic spline to data for the relative scintillation efficiency of Xenon, shown as a yellow line, along with the one sigma contour, shown in blue. The fit uses five knots, shown as red squares, at fixed positions on the x-axis of $5, 10, 25, 50$ and $100$ keVnr. The uncertainty on the extrapolation is reflected in the top and bottom curves of the one sigma blue band. Note that recoil energy refers specifically to nuclear-recoils here.}
\label{fig:Leff}
\end{figure}

Like in \cite{Aprile2011}, we perform a cubic spline interpolation to the four datasets previously mentioned and use five knots, placed at recoil energies of $5, 10, 25, 50$ and $100$ keVnr respectively. The best-fit cubic spline is found by freely varying the y-axis positions of these knots, while minimising the least-squares $\chi^2$ goodness-of-fit parameter between the interpolated spline and the data (see \cite{Manalaysay2010} for a good discussion of the methodology). The result is shown in figure \ref{fig:Leff}, along with the one sigma contour, obtained by looking for the maximum and minimum y-axis positions of the knots which satisfy $\chi^2 < \chi^2_{\mathrm{min}} + 5.89$. 

\begin{figure}[h]
\centering
\includegraphics[scale=0.4]{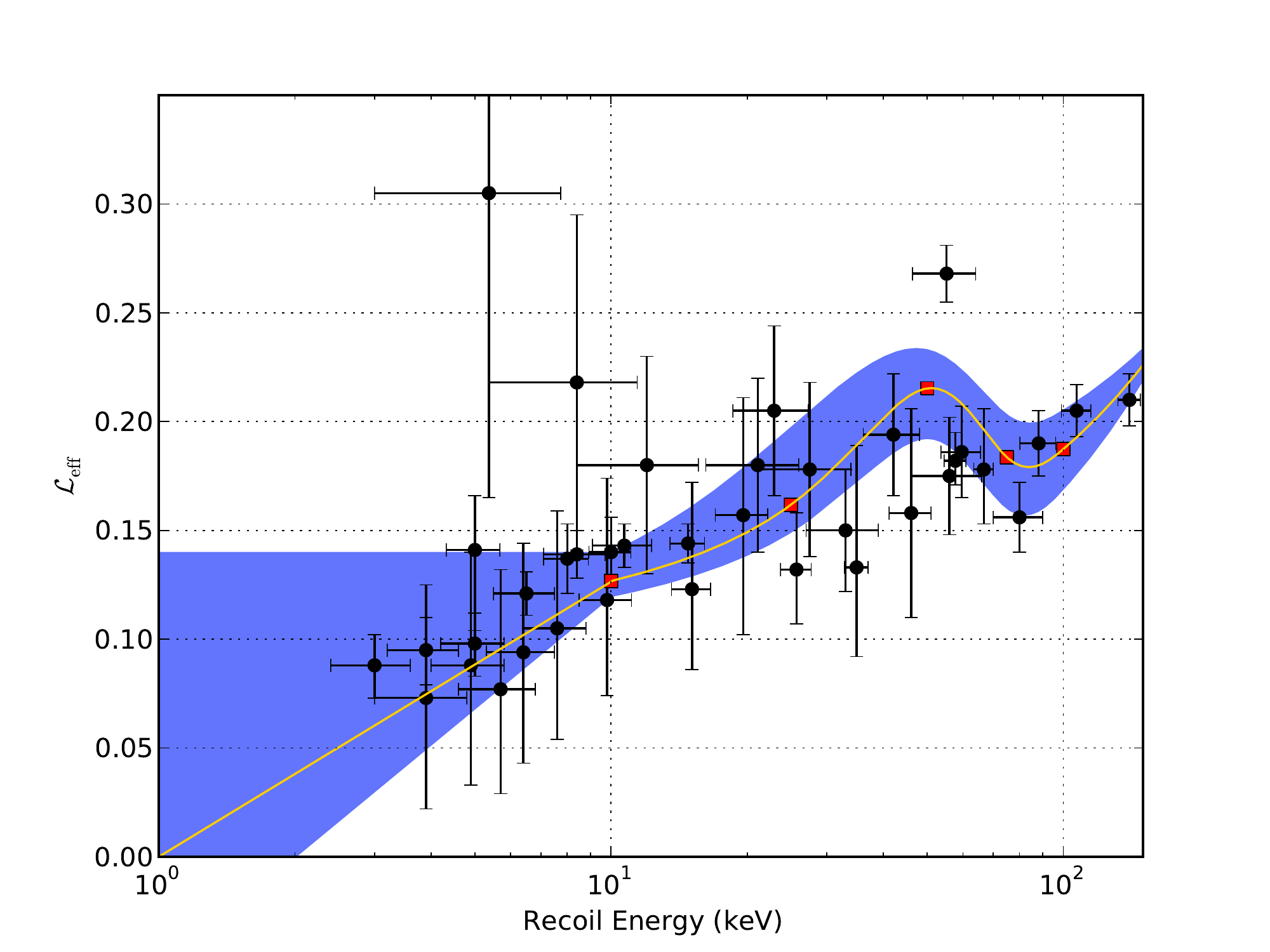}
\caption{A fit of a natural cubic spline to data for the relative scintillation efficiency of Xenon, shown as a yellow line, along with the one sigma contour, shown in blue. The knots used to draw the best-fit spline are shown as red squares, at positions on the x-axis of 10, 25, 50, 75 and 100 keVnr. The uncertainty on the extrapolation is reflected in the top and bottom curves of the one sigma blue band.}
\label{fig:Leff_newknots}
\end{figure}

The choice of the x-positions of these five knots being somewhat arbitrary, we now perform another cubic spline interpolation where we place the knots at $10$, $25$, $50$, $75$ and $100$ $\, \mathrm{keVnr}$. 
The translation of the lowest knot, from $5 \, \mathrm{keVnr}$ to $10 \, \mathrm{keVnr}$, has been performed to illustrate the effect of ignoring the potentially less-reliable data below $10 \, \mathrm{keVnr}$. As can be seen in Fig.~\ref{fig:Leff_newknots}, the greatest change due to the new knot positions ($5 \rightarrow 10$ keVnr and the additional knot at 75 keVnr)  appears to be the enlargement of the errors in the extrapolated region for energies below the first knot. However there are also clear alterations to the interpolation around $75 \, \mathrm{keVnr}$. 

Changing the knots influences the  ${\cal{L}}_{\rm{eff}}$ energy dependence. 
In particular, it changes the shape at high and very low energy. By adding a knot at 75 keVnr, we actually gave some weight to the single point at (55.2,0.268) which  has for effect to drag the curve up around 50 keVnr. Removing the knot at 5 keVnr and instead extrapolating also changes the behaviour of ${\cal{L}}_{\rm{eff}}$  at low energy. In particular, the uncertainties on ${\cal{L}}_{\rm{eff}}$ become larger below 10 keVnr and notably the constant extrapolation moves to higher values of ${\cal{L}}_{\rm{eff}}$.

\subsection{${\cal{L}}_{\rm{eff}}$ extrapolation}

Since there are no data-points below nuclear-recoil energies of $3 \, \mathrm{keVnr}$ there is a great uncertainty on the energy dependence of ${\cal{L}}_{\rm{eff}}$ at low recoil energies. The empirical behaviour found in \cite{Manzur2010} seems to imply that ${\cal{L}}_{\rm{eff}}$ falls down to 0 at low energy in a way which would be consistent with the spline fit of ${\cal{L}}_{\rm{eff}}$ at higher energy. However, \cite{Bezrukov2011} suggests that based on the physics of Xenon recoil and an understanding of both the ionisation yield and scintillation efficiency, ${\cal{L}}_{\rm{eff}}$ should be constant below 10 keVnr. Such an energy behaviour would be supported by \cite{Manalaysay2010} where it is argued that the drop in the scintillation efficiency observed by \cite{Manzur2010} could be due to the drop in sensitivity in the experiment.

Given the lack of data, we will perform an extrapolation of our curves at low energy as in \cite{Aprile2011}. I.e. we adopt either a constant ${\cal{L}}_{\rm{eff}}$ below a certain energy threshold or a drop to 0. For this latter case, we either extend the spline fit to 1 keVnr  or  to 2 keVnr (as in \cite{Aprile2011}). The uncertainty on the extrapolation is reflected in the top and bottom curves of the one sigma blue band in both Figs.\ref{fig:Leff} and \ref{fig:Leff_newknots}. Finally, we also try a sharp cut-off of ${\cal{L}}_{\rm{eff}}$ at low energy for the bottom curve of Fig.\ref{fig:Leff_newknots} in order to obtain the most conservative limit.

\subsection{Robustness of the fit}

Figures \ref{fig:Leff} and \ref{fig:Leff_newknots} show that even with slight modifications in the fitting procedure, the results for $\mathcal{L}_\mathrm{eff}$ as a function of recoil energy can change significantly. In order to check the quality of a certain fit to the data, we employ the \textit{extended critical filter} formalism presented in \cite{Oppermann2011}. This formalism finds a fit to a noisy data set by making use of the error statistics of the data points as well as a Gaussian prior probability distribution for the underlying curve. It is taking into account the possibility of outliers in the data, i.e.\ data points with significantly underestimated error bars. This seems to be beneficial in the case of the $\mathcal{L}_\mathrm{eff}$ measurements due to the wide spread and apparent inconsistency of the different data sets.

Here, we feed the algorithm with different $\mathcal{L}_\mathrm{eff}$-curves as mean for the Gaussian prior. If the prior mean is already a sufficiently good fit to the data set, the result of the extended critical filter procedure will not deviate from it. If, on the other hand, the result of the data filtering differs from the prior mean input, it is a sign that the data prefer a different curve, even though the possibility of individual data points being outliers is accounted for. These outliers are accounted for in the algorithm by the inclusion of a correction factor for the error bar of each data point (see \cite{Oppermann2011} for all technical details). By narrowing the prior probability distribution for these correction factors, we can force the algorithm to take each data point more seriously and thus find out which of the fits is most consistent with the data.

In this way, we study the quality of the two cubic spline fits shown in Figs.~\ref{fig:Leff} and \ref{fig:Leff_newknots}, as well as the $\mathcal{L}_\mathrm{eff}$-curves given by the upper and lower one-sigma contours (i.e. the edges of the blue-shaded regions in Figs.~\ref{fig:Leff} and \ref{fig:Leff_newknots}). Using a reasonably wide prior for the error bar correction factors, we find that all of these curves are consistent with the data, except the top edge of the one-sigma region in Fig.~\ref{fig:Leff}. The exclusion of this one curve might, however, well be due to its behavior at large recoil energies and is likely not to be related to the extrapolation at lowest energies since the top one-sigma curve in Fig.~\ref{fig:Leff_newknots} is not excluded although it is a more extreme extrapolation. Note also that the behavior at recoil energies below $3~\mathrm{keVnr}$ is not constrained by this analysis.

\begin{figure}[h]
    \centering
    \includegraphics[scale=0.4]{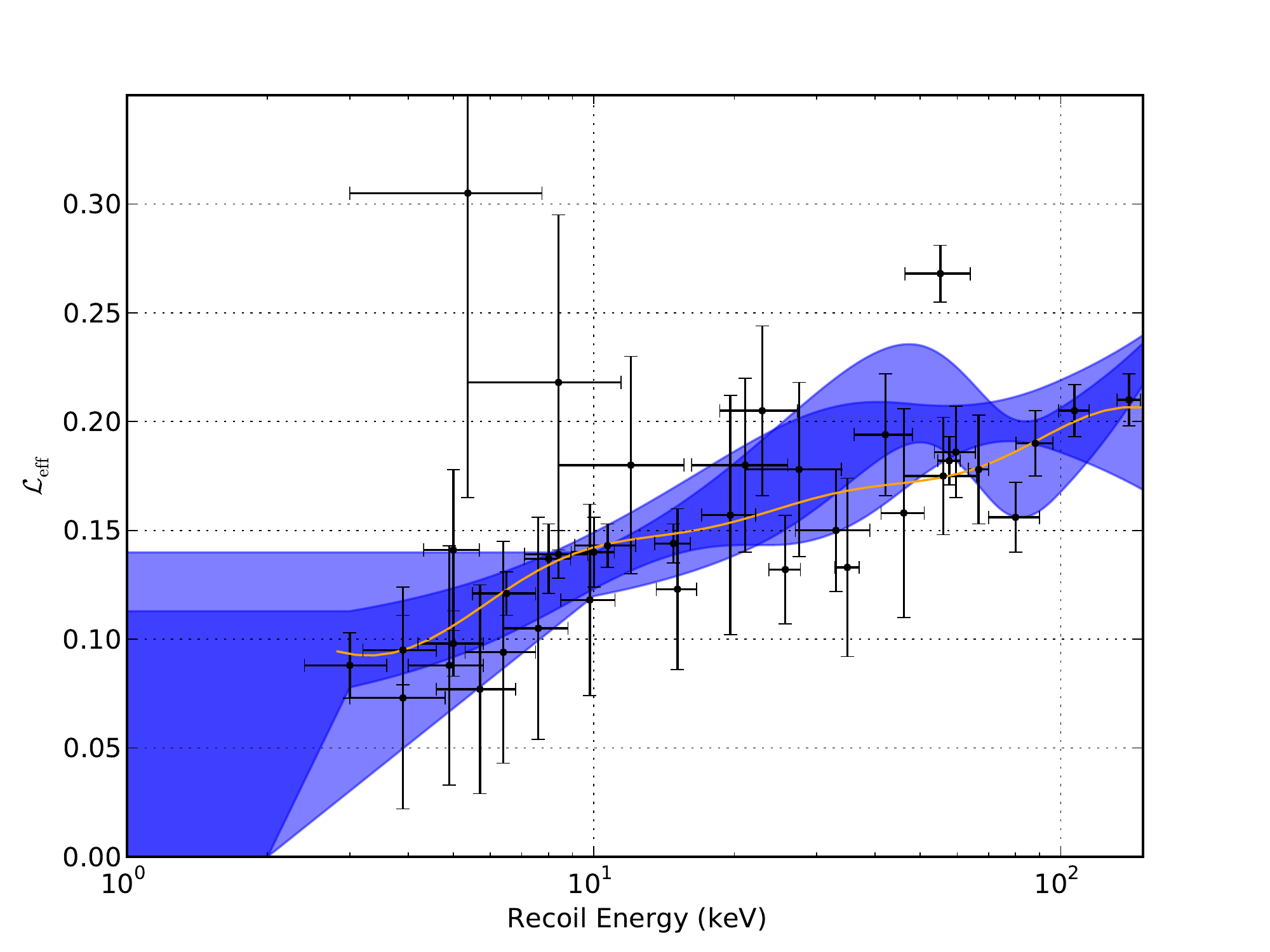}
    \caption{Reconstruction of the calibration curve when using the extended critical filter with a constant prior mean (corresponding to the mean of all data points), shown as a yellow line, along with the one sigma contours around the fits of Figs.~\ref{fig:Leff} and \ref{fig:Leff_newknots}.}
    \label{fig:ECFresult}
\end{figure}

When narrowing the prior for the error bar correction factors to more and more extreme shapes, more curves are successively excluded. It can thus be determined that the central fit in Fig.~\ref{fig:Leff} is the most consistent one with the data. The multitude of $\mathcal{L}_\mathrm{eff}$-curves that is consistent with the present data, however, clearly underlines the importance of studying their influence on the resulting exclusion curve. In fact, yet another fit can be obtained by using a constant curve as prior mean for the extended critical filter and narrowing the prior for the error bar correction factors until deviations from this constant become significant. The resulting curve is shown in Fig.~\ref{fig:ECFresult}, along with the one-sigma contours of the two spline-fits shown in Figs.~\ref{fig:Leff} and \ref{fig:Leff_newknots}.

\section{Exclusion limit  \label{exclu}}

Now that we have determined the uncertainties on ${\cal{L}}_{\rm{eff}}$, we can compute the counting rate of dark matter events expected in the XENON100 detector and deduce an exclusion limit for a given ${\cal{L}}_{\rm{eff}}$. For this purpose, we use a profile Likelihood ratio method and compute p-values for the signal and background, as done in \cite{Aprile2011} after randomly simulating 10000 'mock' data sets based on the XENON100 data published in \cite{Aprile2011}.

\subsection{Counting rate}
The recoil rate (per nucleus) is parameterised in the standard form of \cite{Savage2009},
\begin{equation}
\label{eqn:recoil_rate}
\frac{\mathrm{d}R}{\mathrm{d}E} = \frac{\sigma (q)}{2 \, m \, \mu^2} \ \rho \, \eta (E,t) ,
\end{equation}
where $\sigma$ is the WIMP-nucleus cross-section, $q = \sqrt{2 m_N E}$ is the nuclear recoil momentum (with $m_N$ being the nucleus mass), $m$ is the WIMP mass, $\mu$ is the WIMP-nucleus reduced mass, $\rho$ is the local WIMP density and $\eta (E,t)$ is the WIMP mean speed, given by the expression
\begin{equation}
\label{eqn:mean_speed}
\eta(E,t) = \int_{v_{\mathrm{min}} (E)}^{\infty} \frac{f(v, u_{\mathrm{e}}(t))}{v} \mathrm{d}^3 v \, .
\end{equation}
In the above integral, $u_{\mathrm{e}}(t)$ is the relative velocity between the Earth-based detector and the WIMPs, with time-dependence arising from the motion of the Earth around the Sun, and $v_{\mathrm{min}} (E)$ is the minimum velocity for a WIMP producing a nuclear-recoil of energy $E$. Any astrophysical uncertainties, such as dependance on the galactic WIMP velocity distribution $f(v,u_{\mathrm{e}}(t))$, will arise through this $\eta (E,t)$, and could affect the analysis. Note that the standard halo model is used in the proceeding analysis and  we 
took the average velocity over the year and multiplied the final rate by 100 days (in accordance with \cite{Aprile2011} where the total number of expected events assumes a run of $100$ days with a fiducial volume of $48$ kg).

The WIMP-nucleus cross-section $\sigma$ can be further parameterised (assuming spin-independent interactions, and equal coupling to protons and neutrons) as,
\begin{equation}
\label{eqn:cross_section}
\sigma (E) = \sigma_N \left( \frac{\mu}{\mu_p} \right)^2 A^2 F^2 (E) \, ,
\end{equation}
where $\sigma_N$ is the zero-momentum-transfer cross section for WIMP-nucleon interactions (hereafter any references to cross-section will be to $\sigma_N$), $A$ is the atomic mass, $\mu_p$ is the WIMP-proton reduced mass and $F(E)$ is the nuclear form factor (taken to be of the standard Helm form here) \cite{Lewin1996,Cerdeno2010}.
Equation \eqref{eqn:recoil_rate} can now be used to calculate the signal rate per number of photoelectrons $n$ in the detector:
\begin{equation}
\label{eqn:dRdn}
\frac{\mathrm{d}R}{\mathrm{d}n} = \int_0^{\infty} \mathrm{d}E  \times \frac{\mathrm{d}R}{\mathrm{d}E} \times P(n,\nu(E)) ,
\end{equation}
where $P(n,\nu(E))$ refers to the Poisson distribution $P(n,\nu(E)) = \frac{\nu^n e^{-\nu}}{n!}$ \cite{Aprile2011c}. Here $\nu(E)$ is the expected number of photoelectrons in the detector at an energy $E$ and  is given by the expression
\begin{equation}
\label{eqn:nu}
\nu(E) = E \times \mathcal{L}_{\mathrm{eff}} (E) \times \frac{S_{\mathrm{nr}}}{S_{\mathrm{er}}} \times L_y,  
\end{equation}
where $ \mathcal{L}_{\mathrm{eff}} (E) \ \times L_y$ gives the number of expected scintillation photons from a nuclear recoil per $\mathrm{keVnr}$ of nuclear-recoil energy $E$. 

In order to translate the expected rate per photoelectron $\frac{\mathrm{d}R}{\mathrm{d}n}$ into something comparable with data, one must factor in the finite photomultiplier resolution $\sigma_{\mathrm{PM}} = 0.5 \, \mathrm{PE}$ and knowledge about cut acceptance $\eta_{\mathrm{cuts}}$. This gives an expression for the rate per nuclear-recoil signal ($S1$):
\begin{equation}
\label{eqn:dRdS1}
\frac{\mathrm{d}R}{\mathrm{d}S1} = \sum_{n=1}^{\infty} \frac{\mathrm{d}R}{\mathrm{d}n} \times G(S1;n,\sqrt{n} \sigma_{\mathrm{PM}}) \times \eta_{\mathrm{cuts}} ,
\end{equation}
here $G(S1;n,\sqrt{n} \sigma_{\mathrm{PM}})$ is a Gaussian distribution
\begin{equation}
G(S1;n,\sqrt{n} \sigma_{\mathrm{PM}})  = \frac{1}{\sqrt{2 \pi n \sigma_{\mathrm{PM}}^2}} \exp{\left(-\frac{(S1 - n)^2}{2 n \sigma_{\mathrm{PM}}^2}\right)} .
\end{equation}

It is then possible to calculate the expected number of signal events in the detector $N_s$, by integrating 
\begin{equation}
\label{eqn:N_s}
N_s (m, \sigma) = \int_{S1_{\mathrm{lower}}}^{S1_{\mathrm{upper}}} \frac{\mathrm{d}R}{\mathrm{d}S1} \mathrm{d}S1 .
\end{equation}
In this case the region between $S1_{\mathrm{lower}} = 4$ and $S1_{\mathrm{upper}} = 20$ is considered.

\subsection{Profile Likelihood ratio}
\label{sec:stats}
The analysis described below follows the approach presented in \cite{Aprile2011c} and centres around the application of the profile likelihood method. The analysis takes as input, both theoretical parameters, such as the expected number of signal events $N_s$ for a given WIMP mass $m_{\chi}$ and cross-section $\sigma$, and a set of data-points. 

\subsubsection{statistic test}
The so-called Profile Likelihood ratio for a particular dataset can be expressed as 
\begin{equation}
\label{eqn:lambda}
\lambda(\sigma) = \frac{\mathcal{L}_{max}(\sigma)}{\mathcal{L}_{max}(\hat \sigma)},
\end{equation}
where here $\mathcal{L}_{max}(\sigma)$ refers to the likelihood function maximised with respect to all parameters but $\sigma$, which is held fixed, and $\mathcal{L}_{max}(\hat \sigma)$ refers to the likelihood maximised with respect to all variable parameters\footnote{these are specifically cross-section $\sigma$, galactic escape velocity $v_{\mathrm{esc}}$, total number of background events $N_b$ and the probabilities to be signal and background events $\epsilon_s^j$ and $\epsilon_b^j$} respectively. The quantity $q_{\sigma}$ defined as 
\begin{equation}
\label{eqn:qsigma}
q_{\sigma} = \begin{cases} -2 \mathrm{ln} \lambda (\sigma) &\hat{\sigma} < \sigma \\
0 &\hat{\sigma} > \sigma , \end{cases}
\end{equation}
with $\hat{\sigma}$ being the value of the cross-section which extremises the likelihood function measures the quality of the fit for that particular dataset and dark matter parameters (mass and cross section). 
The larger $q_{\sigma}$ is, the less signal-like these parameters are supposed to be, thus ruling out this particular value of $\sigma$ for a given dark matter mass and dataset. In principle, for every mass one should test the entire range of cross section that one initially considered. Here, however, we only tested values of the cross section which are relatively close to the XENON100 limit by using a step-size of $\log_{10}(\sigma) = 0.02$ in a range defined as $[\sigma_{xenon},10 \ \sigma_{xenon}]$ with $\sigma_{xenon}$ the value of the dark matter-nuclei cross section at the exclusion limit.

The question of the set of data-points that should be considered for the analysis is essential. In the following, we will use both the experimental dataset given in \cite{Aprile2011} and simulated datasets that we will generate using a Monte Carlo. By considering hypothetical alternate XENON100 experiments, represented by randomly simulated datasets (see subsection\ref{datsets}),  one therefore takes into account the random nature of the experimental data. With this in mind, there should be a certain proportion of simulated datasets (the exact value of which depends on the chosen confidence) which provide a better fit than the actual experimental data. 

The  log of the statistical test ($q_{\sigma}$) for a given dataset, dark matter mass and cross section is uniquely determined. We will refer to it as $q_{\sigma}^{obs}$ when using the experimental data set and $q_{\sigma}$ otherwise. To a given simulated data set corresponds a certain value of $q_{\sigma}$ (for fixed dark matter parameters). Thus, one expects a $q_{\sigma}$ distribution, which can be used to define a $p$-value.

The signal and background p-values ($p_s$ and $p_b$ respectively), are defined as:
\begin{eqnarray}
\label{eqn:ps}
p_s &=& \int_{q_{\sigma}^{\mathrm{obs}}}^{\infty} f(q_{\sigma}, H_{\sigma}) \mathrm{d} q_{\sigma} \\
\nonumber\\
\label{eqn:pb}1 - p_b &=&\int_{q_{\sigma}^{\mathrm{obs}}}^{\infty} f(q_{\sigma}, H_{0}) \mathrm{d} q_{\sigma}.
\end{eqnarray}
Here $f(q_{\sigma}, H_{\sigma})$ is the probability density function (pdf) of all the $q_{\sigma}$ values from simulated datasets, under the so-called signal hypothesis $H_{\sigma}$, while $f(q_{\sigma}, H_{0})$ is the pdf for the background hypothesis $H_{0}$ (cf subsection \ref{datsets}) for fixed dark matter parameters. Here $H_{0}$ refers to the electronic recoil events only while $H_{\sigma}$ refers to the electronic and nuclear recoil events.

Under a desired confidence of $90 \%$, one defines the exclusion curve by satisfying the condition that
\begin{equation}
p_s (\sigma) = 0.1 \times (1 - p_b (\sigma))
\label{eqn:ps_pb}
\end{equation}
for each dark matter mass. Incorporating the p-value for the background hypothesis $H_0$ into the calculation of the exclusion curve enables to take into account the possibility that the background can mimic a WIMP discovery signal, by explicitly requiring the chosen cross-section and dark matter mass to fit better to the signal hypothesis $H_{\sigma}$ than $H_0$. This is particularly important given the overabundance of background points compared to signal points for the XENON100 data.

\subsubsection{Likelihood Function}

Our likelihood function is the same as in \cite{Aprile2011c} except that  we do not parameterise the uncertainty in the relative scintillation efficiency so as to make explicit the impact of ${\cal{L}}_{\rm{eff}}$ 
on the exclusion limit.

Like \cite{Aprile2011c},  we make use of bands to discriminate between electronic and nuclear recoils in S1-S2 space, thereby separating the signal from the  background, allowing a more stringent limit to be placed on WIMP mass and interaction cross-section.

The likelihood function ($\mathcal{L}$) is given by the equation below:
\begin{eqnarray}
\mathcal{L} &=& f_v (v_{\mathrm{obs}}, v_{\mathrm{esc}}) \times \nonumber \\ 
&&\Bigg( \prod_{j=1}^{23} P(n^j, (\epsilon_s^j N_s + \epsilon_b^j N_b)) 
\times \nonumber \\ 
&&P(m_b^j,(\epsilon_b^j M_b)) \times P(m_s^j, (\epsilon_s^j M_s)) \Bigg) , \nonumber 
\end{eqnarray}
where here $j$ is run over each of the 23 bands (the bands themselves are shown in figure \ref{fig:sg_dataset}), and $n_j$ is the number of data points in band $j$. 

The first term parameterises uncertainty in the escape velocity $v_{\mathrm{esc}}$, which is treated as a nuisance parameter here. The second term compares, using a Poisson-distribution function, the number of expected data-points in each band $j$ for both signal and background, to the actual number of points $n_j$. Here $\epsilon_b^j$ is the probability for a background event to be in band $j$ and $\epsilon_s^j$ is the equivalent for signal events, as determined from calibration data.
We start by defining $\epsilon_b^j $ as $\epsilon_b^j = m_b^j/M_b$
where $m_b^j$ is the number of electronic recoil data points in band j 
as appearing in the calibration data, $M_b = \sum_j m_b^j$, and then marginalise over it. Similarly, before marginalisation, $\epsilon_s^j = \frac{m_s^j}{M_s}$ where $m_s^j$ is the number of nuclear  recoil data points in band j as appearing in the calibration data $M_s = \sum_j m_s^j$.  The product $\epsilon_s^j  N_s$ represents the expected number of signal events which fall into the nuclear recoil bands (given a particular cross section, WIMP mass and choice of ${\cal{L}}_{\rm{eff}}$).

Since the expected total number of signal events $N_s$ is a function of cross-section $\sigma$, data-points in bands where $\epsilon_s^j > \epsilon_b^j$ will have the greatest effect on the best-fit $\sigma$ value for a particular dataset. In particular the analysis is very sensitive to any data-points appearing in the lower bands (see figures \ref{fig:sg_dataset} and \ref{fig:cal_contours}), where electronic recoil/background events are unlikely to occur.

Finally the last two terms parameterise the uncertainty in probabilities $\epsilon_s^j$ and $\epsilon_b^j$, due to the expected Poisson-variance of the number of $^{241}$AmBe and $^{60}$Co calibration data-points in each band, $m_s^j$ and $m_b^j$ (here $M_s$ and $M_b$ are the total numbers of nuclear and electronic recoil points respectively for the calibration data).

\subsubsection{Dataset Simulation \label{datsets}}
\label{sec:dataset_sim}
Since simulated datasets play a vital role in determining the pdfs $f(q_{\sigma}, H_{\sigma})$ and $f(q_{\sigma}, H_{0})$, it is important to discuss their method of generation, and the uncertainties involved.

As discussed in the previous section, by defining an exclusion curve at $90 \%$ (or any value different from $100 \%$) , the naturally random nature of the experiment is taken into account. Since the XENON100 experiment can only ever be performed once, it is possible that any observation, or non-observation, of possible signal data-points could be due, wholly or in-part, to  statistical fluctuations. One seeks to improve this possibility by positing hypothetical alternate XENON100 experiments, which differ only in their sampling of the statistics. Practically these alternate data-sets are represented by simulations, based on the actual experimental data \cite{Aprile2011}.

Since the simulated data should attempt to mimic that obtained by the experiment, the data-points must be arranged on the S1-S2 plane, where S1 is equal to the number of photoelectrons observed for a particular event, and S2 represents the ionisation yield. This can be achieved with knowledge of the expected distribution of nuclear-recoils, associated with WIMP signal events, and electronic-recoils, associated with background events, in S1-S2 space. Such information is contained in the calibration data obtained by the XENON100 experiment, who used a $^{60}$Co source for samples of electronic recoils and an $^{241}$AmBe source for nuclear recoils. 

\begin{figure}[h]
\centering
\includegraphics[scale=0.4]{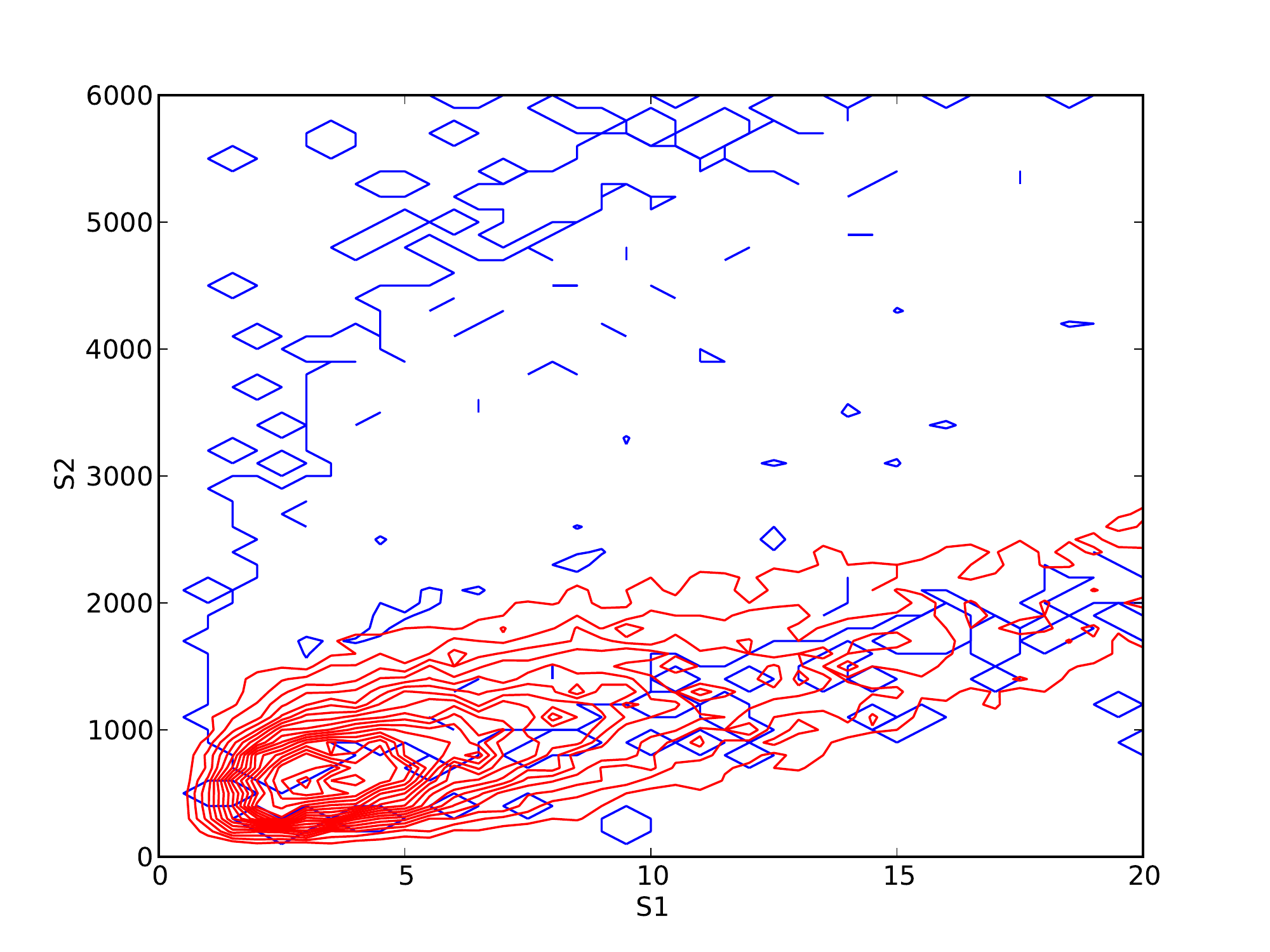}
\caption{Contour plots of calibration data on the S1-S2 plane, showing the distribution of nuclear recoils in red and electronic recoils in blue.}
\label{fig:cal_contours}
\end{figure}

In practice this calibration data is binned and normalised, to give a pdf for signal and background events, shown in figure \ref{fig:cal_contours}. Given a desired number of signal (nuclear-recoil) and background (electronic-recoil) events, a simple Monte Carlo algorithm can be used to generate simulated datasets.

It is possible to extend the simulation algorithm further, and improve the accuracy of the simulated datasets, by incorporating information about the WIMP energy spectrum, equation \eqref{eqn:dRdS1}, into the determination of the nuclear-recoil data-points\footnote{Such an extension is not necessary for the background points, as electronic recoils have been shown to have a flat spectrum at the energies considered here \cite{Aprile2011b}}. However, since there are very few candidate signal points seen in the data, and to avoid possible problems with bias, this extension has not been incorporated into the current analysis. Even so, the uncertainty in exactly how to simulate datasets most accurately will contribute a source of uncertainty to the final exclusion curve.

The above method was used to generate $10000$ simulated datasets, with an expected number of $2$ signal events (nuclear-recoils) and $534$ background events (electronic-recoils), between $S1 = 4$ and $S1 = 20$, as seen in the data obtained by the run of the XENON100 experiment \cite{Aprile2011} after 100 live days of data-taking \footnote{As an aside, by extending the S1 region down to zero, discrimination between signal and background points would be easier in principle, due to the clear difference in shape between the probability contours of nuclear and electronic recoils at low S1, as seen in figure \ref{fig:cal_contours}. However the increase in sensitivity is granted at the cost of greater susceptibility to systematics, especially the $\mathcal{L}_{\mathrm{eff}}$ uncertainty.}. In addition to these 'signal + background' datasets, so-called 'background-only' datasets were generated, with an expected number of $536$ background events and no signal events. 

\begin{figure}[h]
\centering
\includegraphics[scale=0.4]{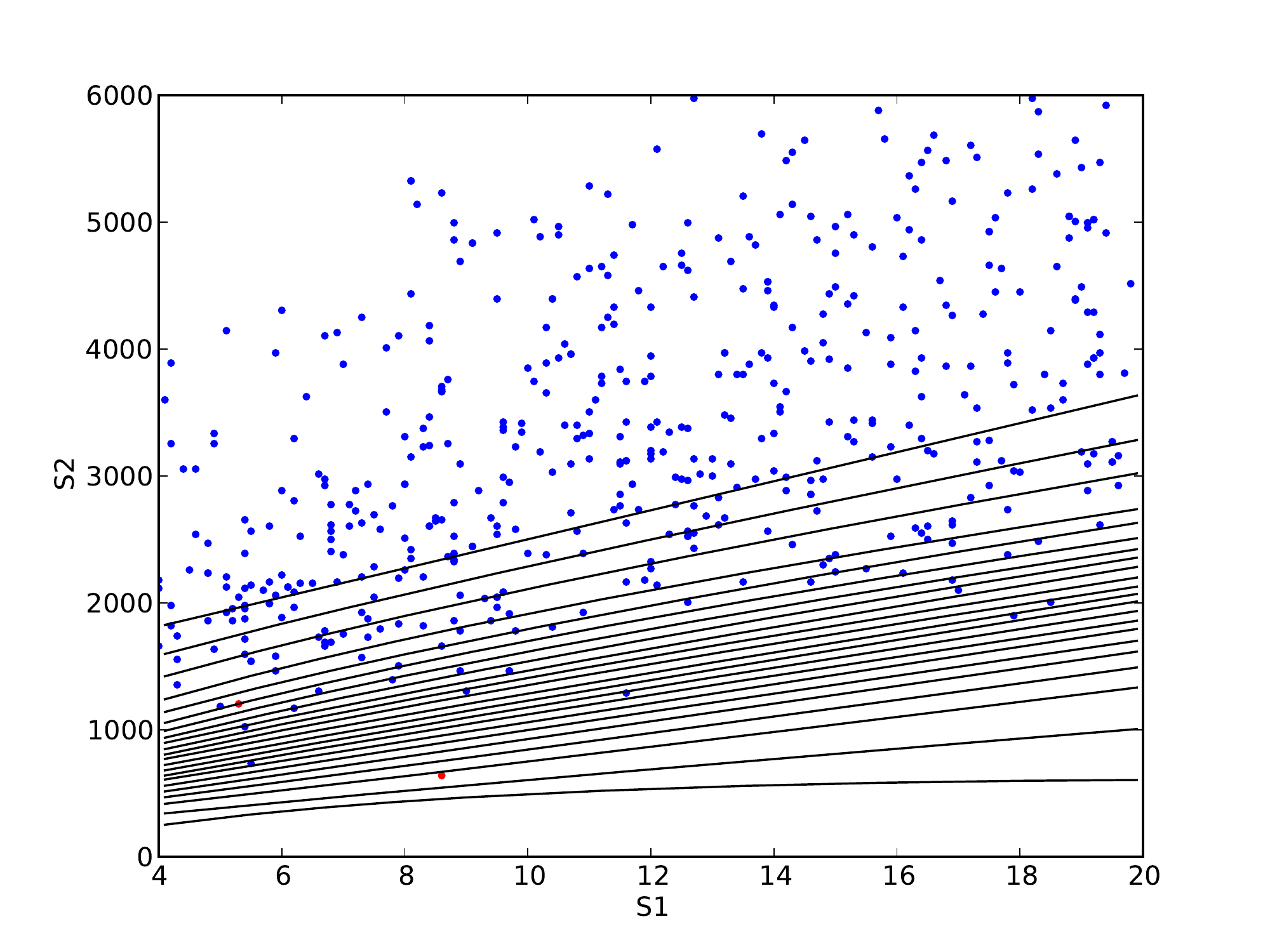}
\caption{An example of a simulated dataset, with two nuclear-recoil (signal) events, shown in red. The rest of the points are electronic-recoil (background), shown in blue. The black lines divide the S1-S2 plane into the bands used for the analysis.}
\label{fig:sg_dataset}
\end{figure}

A plot of one such signal+background simulated dataset is shown in figure \ref{fig:sg_dataset}. The analysis itself is blind to whether a point was generated as a nuclear or electronic recoil, however the fitting of the cross-section is more sensitive to the lower bands, where fewer background events are expected. Due to the abundance of electronic-recoil events compared to nuclear-recoils, determining which points are due to which is a difficult challenge, and so a clearer discrimination between signal and background only arises statistically when considering many such datasets, motivating the choice of a confidence limit other than $100 \%$ for the XENON100 limit. Hence, even with high statistics, the ability of the analytical tools to discriminate signal from background is limited, contributing a natural source of error to any determination of the best-fit values of the cross-section and number of background events, and so ultimately to the final exclusion curve.

Values of $q_{\sigma}$ were calculated for each dataset under the prescription of section \ref{sec:stats}. The signal+background and background-only $q_{\sigma}$ values were then binned  separately into two normalised histograms (for each value of cross-section and WIMP mass), to give the pdfs $f(q_{\sigma}, H_{\sigma})$ and $f(q_{\sigma}, H_{0})$ respectively. In this way, the signal+background datasets represent the signal hypothesis $H_{\sigma}$, as they are generated under the assumption that the two candidates-signal events seen in the XENON100 data are in fact due to nuclear recoils. Conversely the background-only datasets take these points to be due to background electronic-recoils, thereby coming under the background hypothesis.

An example of $f(q_{\sigma}, H_{\sigma})$, for a specific WIMP mass and cross-section, is shown in figure \ref{fig:qsigmapdf7}. Note that, due to Wilks' theorem one expects the pdf to approach a $\chi^2$ distribution as the number of sampled datasets increases, a trend which is indeed observed.
\begin{figure}[htb]
\centering
\includegraphics[scale=0.4]{chi2_pdf_2_-460_m=7_t=0.pdf}
\caption{The probability distribution function of $q_{\sigma}$ under the signal hypothesis, for a WIMP mass of $m=7 \, \mathrm{GeV}$ and $\sigma_N = 2.51 \times 10^{-41} \mathrm{cm}^2$. The pdf has been constructed using $10000$ simulated datasets, assuming $2$ expected signal events and $534$ expected background events. The actual points of the pdf are shown as bars, while the best-fit $\chi^2$ distribution is shown as a black line. The blue dashed line indicates the value of $q_{\sigma}$ for the real experimental dataset.}
\label{fig:qsigmapdf7}
\end{figure}
Each pdf is fitted with an analytical $\chi^2$ function, to speed up computation and to avoid susceptibility to statistical fluctuations at higher values of $q_{\sigma}$. The value of $\sigma$ which satisfies equation \eqref{eqn:ps_pb} is sensitive to this fit, for both $f(q_{\sigma}, H_{\sigma})$ and $f(q_{\sigma}, H_{0})$, and so any uncertainty in the best-fit $\chi^2$ function will contribute to the uncertainty in the final exclusion curve.

\section{Results \label{sec:results}}
\subsection{Systematic Error in XENON100 Exclusion Curves}
\begin{figure}[h]
\centering
\includegraphics[scale=0.4]{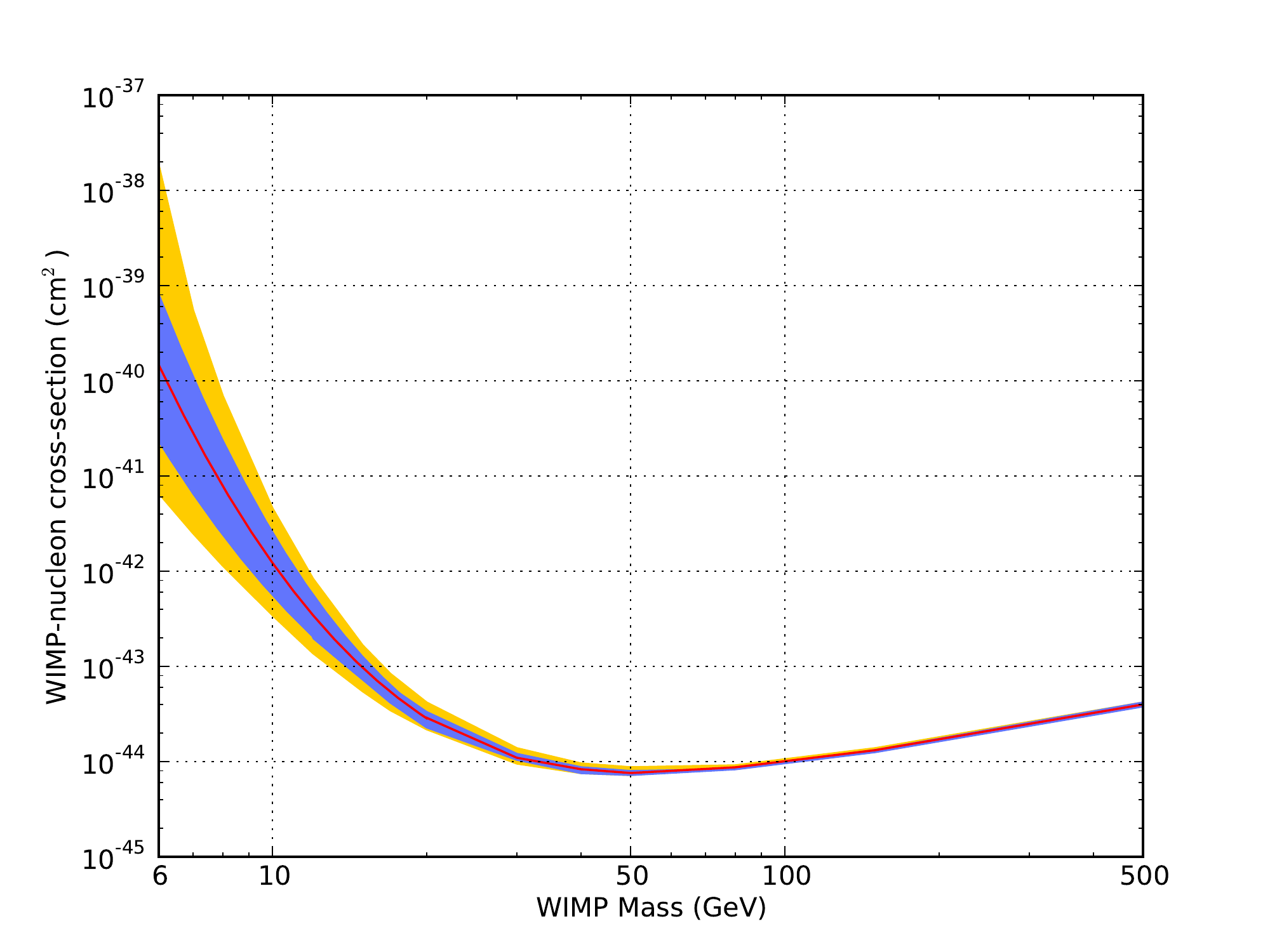}
\caption{The XENON100 exclusion curve using the best-fit cubic-spline for $\mathcal{L}_{\mathrm{eff}}$ from figure \ref{fig:Leff}, shown in red, along with the one sigma systematic uncertainty due to $\mathcal{L}_{\mathrm{eff}}$ from the fit of figure \ref{fig:Leff}, in blue, and the one sigma uncertainty from the fit of figure \ref{fig:Leff_newknots}, in yellow. Note that all exclusion curves have a natural uncertainty of $\log_{10} \sigma = \pm 0.02$.}
\label{fig:Onesigmacurve}
\end{figure}

The profile likelihood analysis has been performed using the best cubic spline fit to $\mathcal{L}_{\mathrm{eff}}$, along with the top and bottom edges of the one-sigma bands from figures \ref{fig:Leff} and \ref{fig:Leff_newknots}. The resulting exclusion curves are shown in figure \ref{fig:Onesigmacurve}. 

Clearly the systematic uncertainty due to the relative scintillation efficiency is appreciably large for WIMP masses below $10 \, \mathrm{GeV}$, with the majority of the variation arising from the extrapolation-uncertainty for $\mathcal{L}_{\mathrm{eff}}$ at low nuclear-recoil energies. The lower edges of the uncertainty bounds on figure \ref{fig:Onesigmacurve} correspond to the upper edges of the one-sigma regions on figures \ref{fig:Leff} and \ref{fig:Leff_newknots}, hence a flat extrapolation of ${\cal{L}}_{\rm{eff}}$ at low-energies tends to result in a stronger XENON100 exclusion limit.

\begin{figure}[h]
\centering
\includegraphics[scale=0.4]{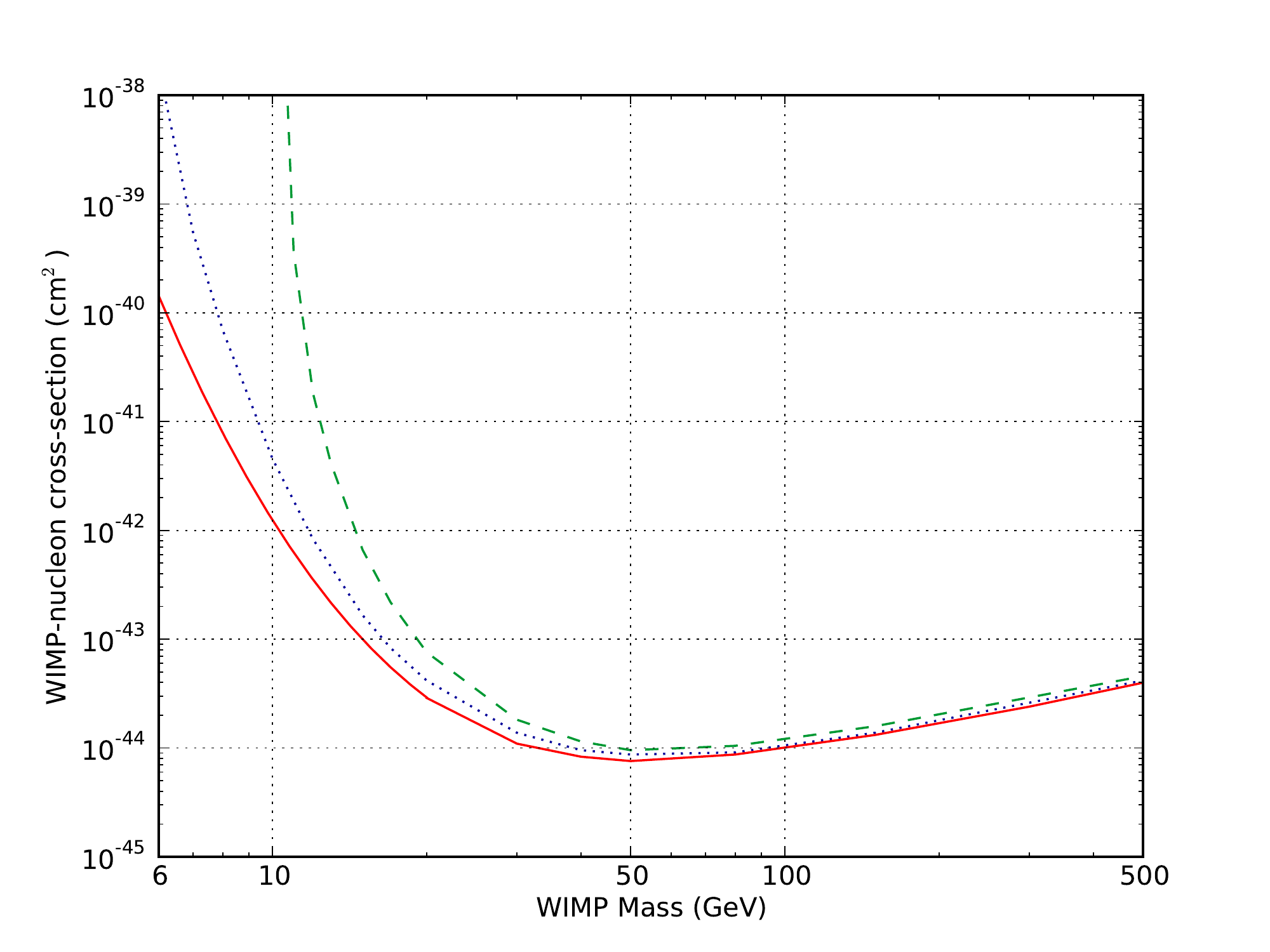}
\caption{The XENON100 exclusion curve as shown above with also the expectations if ${\cal{L}}_{\rm{eff}}$ has a sharp cut-off below 10 keV. The red plain curve is the mean fit; the blue dotted curve represents the 1-sigma contour corresponding to the bottom curve in Fig.~\ref{fig:Leff_newknots} and the green dashed line represents the exclusion curve associated with a ${\cal{L}}_{\rm{eff}}$ function with a sharp cut-off below 10 keV.}
\label{fig:Onesigmacurve_withcutoff}
\end{figure}

The upper curve in Fig.\ref{fig:Onesigmacurve_withcutoff} represents the most conservative exclusion limit that one can derive from the present ${\cal{L}}_{\rm{eff}}$ data. This illustrates that the conclusion from the XENON100 collaboration is fairly robust above $\sim$20 GeV and uncertainties at large mass are really small, as claimed by \cite{Aprile2011}.

In comparison to similar works such as \cite{Savage2011}, which analysed the first set of XENON100 data without using the Profile Likelihood method, our conclusions are similar, but not identical, in the low-mass region of parameter space, for the case where their cut-off of events at $S1 = 1$ was relaxed. However, in the current analysis it is possible that the use of the Profile Likelihood method has changed the size of the systematic $\mathcal{L}_{\mathrm{eff}}$ errors for low-masses due to increased sensitivity, in addition to the different choices of spline interpolation for $\mathcal{L}_{\mathrm{eff}}$ itself, compared with \cite{Savage2011}.

It should be noted that there are a variety of uncertainties affecting the exclusion curves from the analysis, primarily the uncertainty in fitting the $\chi^2$ distribution to the pdfs of $q_{\sigma}$, but also the flexibility in the actual method of dataset simulation, and the overlap of background and signal regions on the S1-S2 plane (as discussed in section \ref{sec:dataset_sim}).

\subsection{Discussion and Implications}
\begin{figure}[h]
\centering
\includegraphics[scale=0.4]{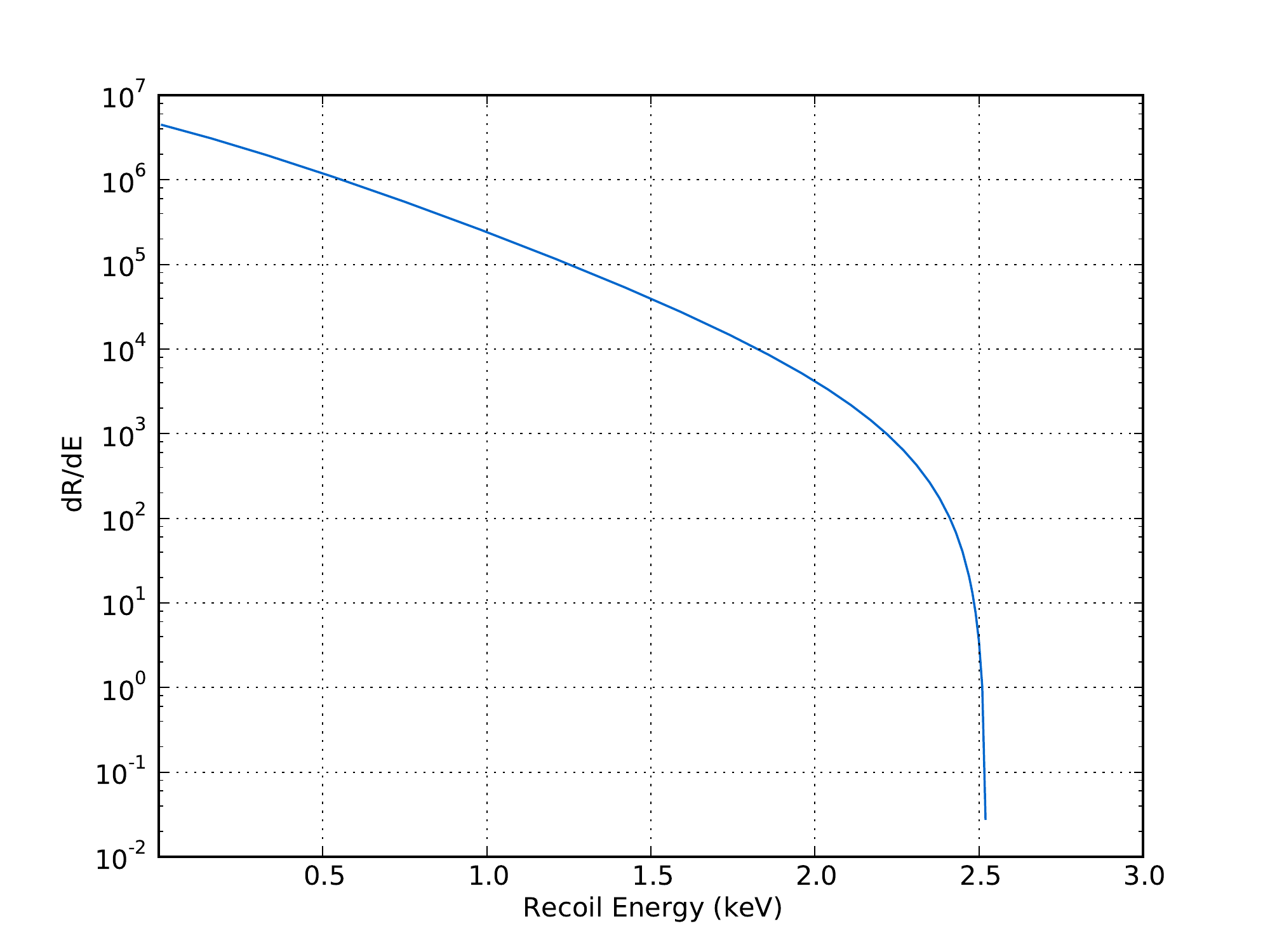}
\caption{The recoil spectrum for a WIMP of mass $5 \, \mathrm{GeV}$ and cross-section of $\sigma_N = 10^{-39} \mathrm{cm}^2$. From zero to $2.5 \, \mathrm{keVnr}$ the differential reaction rate $\frac{\mathrm{d}R}{\mathrm{d}E}$ changes by many orders of magnitude.}
\label{fig:dRdE_m=5}
\end{figure}

The relative size of the variation of the XENON100 exclusion curve with $\mathcal{L}_{\mathrm{eff}}$ at low masses compared to that at high masses, can be understood in terms of the WIMP recoil spectrum, an example of which is shown in figure \ref{fig:dRdE_m=5}, and equations \eqref{eqn:dRdn} to \eqref{eqn:dRdS1}. 

The equation for $\frac{\mathrm{d}R}{\mathrm{d}n}$ \eqref{eqn:dRdn} has two terms in the integrand: the WIMP recoil spectrum $\frac{\mathrm{d}R}{\mathrm{d}E}$ and a Poisson distribution. The Poisson term is peaked at a particular value of energy, which increases for larger numbers of photoelectrons $n$. Hence for a certain value of $n$, there will be a region along the energy axis, of $\frac{\mathrm{d}R}{\mathrm{d}E}$, which contributes most to the integral. By changing the functional form of $\mathcal{L}_{\mathrm{eff}}$, the value of the nuclear recoil energy at which the Poisson distribution peaks will change (by approximately $1 \, \mathrm{keVnr}$ for the different parameterisations considered here). Hence for a particular value of $n$, the integral of equation \eqref{eqn:dRdn} will now receive dominant contributions from different areas of $\frac{\mathrm{d}R}{\mathrm{d}E}$ along the energy axis, when $\mathcal{L}_{\mathrm{eff}}$ is altered.

To explain the different variation in the final exclusion curve due to $\mathcal{L}_{\mathrm{eff}}$ seen at large and small WIMP masses, one must compare recoil spectra. For low masses, the recoil spectrum changes rapidly at low energies (see e.g. figure \ref{fig:dRdE_m=5}), before falling off at a few $\mathrm{keVnr}$. However, at larger masses, the spectrum is largely constant until much higher energies before falling off; in the case of $m_{\chi} = 500 \, \mathrm{GeV}$ the cut-off is at approximately $100 \, \mathrm{keVnr}$. Clearly, the greatest change of $\frac{\mathrm{d}R}{\mathrm{d}n}$ with $\mathcal{L}_{\mathrm{eff}}$ will be seen for values of $n$ where the Poisson term of equation \eqref{eqn:dRdn} is peaked at recoil energies where $\frac{\mathrm{d}R}{\mathrm{d}E}$ varies most rapidly. Since the peak of the Poisson in energy is at larger values for higher $n$, the largest variation in $\frac{\mathrm{d}R}{\mathrm{d}n}$ will be seen for higher masses at high $n$, while for lower masses it will be seen predominantly at low values of $n$.

Finally, equation \eqref{eqn:dRdS1} contains a sum of $\frac{\mathrm{d}R}{\mathrm{d}n}$ over all $n$. For low WIMP masses, the lower values of $n$, where the greatest variation due to $L_{\mathrm{eff}}$ occurs, dominate over the terms with larger $n$. Conversely for larger masses, all values of $n$ contribute terms of the same order to  \eqref{eqn:dRdS1}, hence there will be no dramatic change in the final result, due to the relative scintillation efficiency.

Additionally, in the case of the forms of $\mathcal{L}_{\mathrm{eff}}$ from figures \ref{fig:Leff} and \ref{fig:Leff_newknots}, the systematic uncertainty of the exclusion curve for low WIMP masses is further amplified, relative to the higher masses, due to the larger uncertainties of $\mathcal{L}_{\mathrm{eff}}$ at low energies.


\section{Conclusion \label{sec:conclusion}}

In this paper we have assessed the uncertainties on the XENON100 exclusion curve \cite{Aprile2011} due to the lack of knowledge about the low energy behaviour of the scintillation efficiency of Liquid Xenon detector. Our analysis is motivated by the existence of low mass dark matter scenarios (below 10 GeV)  which lie close to the published XENON100 limit (see for example \cite{Vasquez:2010ru,AlbornozVasquez:2011yq,AlbornozVasquez:2011js}).

The use of a profile likelihood analysis (in which uncertainties on $\mathcal{L}_{\mathrm{eff}}$ were profiled out) enabled the XENON100 collaboration to obtain an exclusion limit which is free from large  uncertainties \cite{Aprile2011}. In particular, the limit on low mass WIMPs seems very precise while one would expect to recover at least the 1-sigma uncertainty band which accounts for the lack of determination of ${\cal{L}}_{\rm{eff}}$ below 3 keVnr.


In order to understand this behaviour, we have performed  
a similar profile likelihood analysis but did not consider ${\cal{L}}_{\rm{eff}}$ as a nuisance parameter. Instead ${\cal{L}}_{\rm{eff}}$ is defined directly from the fits to the data. We show that the exclusion limit obtained by the XENON100 collaboration at high energy is very robust. The uncertainties on the exclusion limit due to ${\cal{L}}_{\rm{eff}}$ are very small and all our exclusion curves are very similar to the XENON100 exclusion limit. Such a conclusion was not necessarily obvious since different types of interpolation of ${\cal{L}}_{\rm{eff}}$  data give different behaviours for ${\cal{L}}_{\rm{eff}}$, even at high energy. However this can be understood from our robustness of the fit analysis which shows that all spline fits  at high energy are equally good. Hence heavy dark matter scenarios close to the XENON100 limits will not be affected by a better determination of ${\cal{L}}_{\rm{eff}}$ at high recoil energies. This implies that dark matter scenarios just above the XENON100 limit are probably excluded indeed (provided that other sources of uncertainties which are not accounted for in this analysis are not too large) and, those just below, very close to be ruled out.  

At low recoil energies\footnote{Our analysis of the robustness of the fit shows that the data still prefer a smooth cut-off to zero for ${\cal{L}}_{\rm{eff}}$ below 3 keVnr, even though a different behaviour has been suggested from theoretical arguments \cite{Bezrukov2011}.}, our results (cf Fig.\ref{fig:Onesigmacurve}) show that the mean value of ${\cal{L}}_{\rm{eff}}$ (which is in agreement with the mean ${\cal{L}}_{\rm{eff}}$ curve considered by the XENON100 collaboration) gives an exclusion limit that is similar to (although stronger than) the XENON100 limit but a more extreme behaviour of ${\cal{L}}_{\rm{eff}}$ at low energy (cf the flat extrapolation or a sharp cut-off below 3 keVnr for example) leads to a very different exclusion limit. Should new data favour such a type of behaviour for ${\cal{L}}_{\rm{eff}}$ at very low energy with unprecedented precision, the exclusion limit would be different from that presented by the XENON100 collaboration, even though such a behaviour was included in the 1- and 2-sigma contours considered for ${\cal{L}}_{\rm{eff}}$ by the XENON100 collaboration. 

Perhaps a reason for this discrepancy is the Gaussian Likelihood term which was assumed for ${\cal{L}}_{\rm{eff}}$ in the XENON100 analysis (and which was centred on the mean value of ${\cal{L}}_{\rm{eff}}$) together with flat priors. Combined with the other Likelihood terms which describe the uncertainties from the analysis itself, it may be that the XENON100 maximum likelihood analysis over-fits
${\cal{L}}_{\rm{eff}}$ and thereby suppresses alternative possibilities for this
crucial quantity, leading to over confidence in excluding light DM
scenarios.

In any case, since various ${\cal{L}}_{\rm{eff}}$ fits give different exclusion limits, it seems more conservative to "track" the effect of different ${\cal{L}}_{\rm{eff}}$ energy behaviour at low energy on the exclusion curve and define (frequentists) confidence intervals by marginalising over the different fits to ${\cal{L}}_{\rm{eff}}$ data. With this method, one is in principle ensured not to bias the analysis towards the present best fit curve of ${\cal{L}}_{\rm{eff}}$ since it may in fact not be the correct function to consider given the lack of data in this energy region. Finally, we note that a flat behaviour of ${\cal{L}}_{\rm{eff}}$ (as suggested by \cite{Bezrukov2011}) at low recoil energy would actually set a stronger exclusion limit. Hence the need for new measurements of ${\cal{L}}_{\rm{eff}}$ at low energy.

Our analysis was based on a likelihood chosen so as to keep
the current analysis as close as possible to the one done by the
XENON100 collaboration \cite{Aprile2011,Aprile2011b,Aprile2011c}, however it should be possible
to improve the method in identifying WIMP signals over background. If the S1-S2 plane were to be divided into a grid of bins, then, on top of the knowledge of the actual structure of the set of data-points in this space from calibration data, one could use additional knowledge of the recoil spectra for background and signal. Such a spectrum is already known along the S1-axis for WIMPs (equation \eqref{eqn:dRdS1}), and is distinct to that from electronic recoils, which is constant for low energies \cite{Aprile2011b}. This would provide an additional method of discrimination, which would be especially important if a future run of the XENON100 experiment were to claim a discovery signal, but is difficult without more knowledge of the experimental set-up. Additionally, it should be possible, in principle, to test the efficiency of a particular likelihood function in reconstructing theoretical parameters such as the cross-section  from given simulated datasets, since the parameters used to generate them are known (see \cite{Strege2012} for a similar discussion).

\section{Acknowledgments}
We would like to thank Michael Schmidt and Hendrik Hoeth for their precious help. JHD is supported by a STFC studentship.


\end{document}